# Asymmetric Cherenkov Emission in a Topological Plasmonic Waveguide

Filipa R. Prudêncio[1,2]* and Mário G. Silveirinha[1]

[1] *University of Lisbon – Instituto Superior Técnico and Instituto de Telecomunicações, Avenida Rovisco Pais 1, 1049-001 Lisboa, Portugal*

[2]*Instituto Universitário de Lisboa (ISCTE-IUL), Lisboa, Portugal*

## Abstract

Here, we investigate the Cherenkov emission by an array of moving electric charges in the vicinity of a topologically nontrivial gyrotropic material. It is shown that the nonreciprocal material response may result in a robustly asymmetric Cherenkov emission, such that the spectrum of the emitted radiation and the stopping power depend strongly on the sign of the particle velocity. It is demonstrated that the main emission channels are determined by the unidirectional edge states supported by the topological material. We consider as examples both magnetized plasmas and Weyl semi-metals. The latter may exhibit a spontaneous nonreciprocal response without a biasing magnetic field.

---

* Corresponding author: filipa.prudencio@lx.it.pt

# I. Introduction

The use of topological methods in photonics opened up a myriad of novel photonic platforms with fascinating properties [1]-[8]. In particular, Chern topological insulators are nonreciprocal structures (e.g., photonic crystals) characterized by a (nontrivial) topological invariant called the Chern number [1, 2]. The topological invariant depends on the material band structure [1] and determines the net number of topologically protected unidirectional edge states supported by an interface of the material and another "mirror" with a trivial topology and a common band gap [1, 2]. The edge states are protected against backscattering and this unique and singular property has inspired the development of many novel photonic platforms insensitive to imperfections and to back-reflections [1]-[8]. Some reciprocal material platforms with a bianisotropic response also have a topological classification [9]-[12]. Moreover, the ideas of topological photonics can be extended to a continuum with no intrinsic periodicity [6, 7] and gyrotropic media are generically topologically nontrivial. Recently, it was shown that the photonic Chern number has a precise physical meaning: it can be understood as the *quantum* of the thermally-generated light-angular momentum in a closed cavity filled with the photonic insulator [13].

Typically, nontrivial topological materials require some external biasing to break the time-reversal symmetry [1, 2], e.g., a magnetic bias. Remarkably, it was recently predicted that a new class of electronic topological materials called Weyl semi-metals [14]-[17] may exhibit a spontaneous nonreciprocal response without any external magnetic field due to magnetic order. Weyl semi-metals have unusual properties, for example, the electrons in a fully three-dimensional solid are described by the massless relativistic Dirac equation [14, 15].

The Cherenkov effect [18]-[20] occurs when a charged particle moves inside or nearby a transparent medium with a velocity *v* larger than the phase velocity of light in the medium. In recent years, there has been a renewed interest in this fundamental effect, largely motivated by the discovery of photonic crystals and metamaterials and by the fact that such structures can lead to exotic forms of Cherenkov radiation [21]-[23]. In particular, Veselago theoretically predicted in 1967 that a moving particle in a left-handed material produces both forward and reversed Cherenkov radiation [21]. More recently, the Cherenkov effect was also investigated in plasmonic platforms and in metamaterials with a plasmonic-type hyperbolic response [24]-[30]. The Cherenkov radiation has relevant applications in free-electron lasers, biomedicine, particle detection and nanoscale light sources [31].

In this article, we investigate the role of a nontrivial topology in the context of the Cherenkov radiation. We consider a beam of electric charges moving in a vacuum in the vicinity of *i*) a gyrotropic slab (e.g., a magnetized plasma) and *ii*) a Weyl semi-metal characterized by a spontaneous nonreciprocal response. We find that the unidirectional plasmon edge waves are the main radiation channels and lead to strongly asymmetric Cherenkov radiation with spectrum depending on the sign of the electrons velocity. Moreover, the studied platforms may behave as "Cherenkov diodes" such that the amount of emitted radiation depends on the sign of the velocity, which may useful in detectors where it is important to discriminate the sign of the particle velocity.

The article is organized as follows. In Section II we present the theoretical formalism for a "pencil beam" of charges. In Section III we study the Cherenkov emission for the cases where the charged particles move in the vicinity of a magnetized plasma and a Weyl semi-metal. In Section IV some of our findings are generalized to the case of a "point-like" beam and the stopping power is analytically calculated. Finally, in Section V the main conclusions are drawn.

## II. Theory for a pencil of moving charges

For simplicity, in a first stage we suppose that the relevant beam of charges is shaped in the form of a "pencil", i.e., it corresponds to a linear array of charges moving in a vacuum near the topological material. The charges move along the $x$-direction with a constant velocity $v$ at a distance $d$ from a gyrotropic-material occupying the negative $y$ half-space (Fig. 1a). The current density describing this pencil beam is $\mathbf{j}_e(x,y,t) = -en_z v \delta(y-y_0)\delta(x-vt)\hat{\mathbf{x}}$, where $-e$ is the electron charge, $n_z$ is the number of charges per unit of length along the $z$ direction, and $y_0 = d$. The outlined problem is effectively two-dimensional which eases the analytical developments. In Section IV, we generalize some of our results to the fully 3D scenario.

The Cherenkov radiation emitted by the linear array of charges is determined by the Maxwell equations, $\partial \mathbf{B}/\partial t + \nabla \times \mathbf{E} = 0$ and $\nabla \times \mathbf{H} - \partial \mathbf{D}/\partial t = \mathbf{j}_e$. For a pencil beam, the fields are of the form $H = H_z(x,y,t)\hat{\mathbf{z}}$ and $\mathbf{E} = E_x(x,y,t)\hat{\mathbf{x}} + E_y(x,y,t)\hat{\mathbf{y}}$. It is convenient to work in the frequency (Fourier transform) domain. The Fourier transform of the electric current density ($\mathbf{j}_e(x,y,\omega) = \int_{-\infty}^{+\infty} \mathbf{j}_e(x,y,t)e^{+i\omega t}dt$) is $\mathbf{j}_e(x,y,\omega) = -en_z \operatorname{sgn}(v)\delta(y-y_0)e^{ik_x x}\hat{\mathbf{x}}$ where $k_x = \omega/v$ and $y_0 = d$.

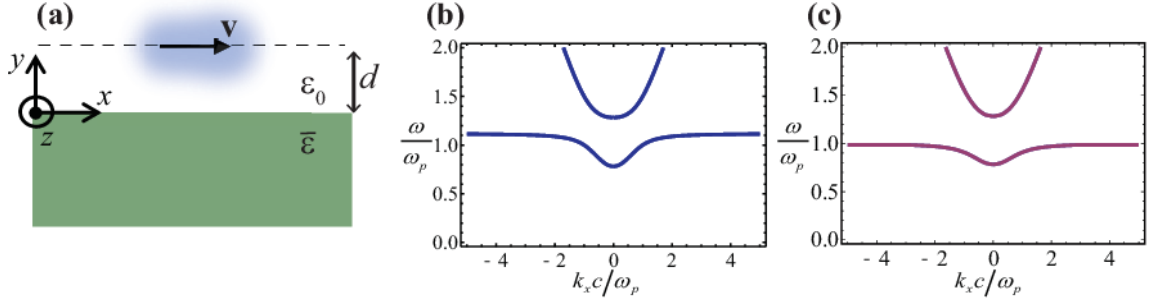

**Fig. 1 (a)** A linear array of electric charges moves with a constant velocity *v* at a distance *d* from either a (i) gyrotropic material (biased along the *z*-direction) or a Weyl semi-metal. **(b)** Band diagram of the natural plane-wave modes for a bulk magnetized plasma. **(c)** Band diagram of the natural plane-wave modes for a bulk Weyl semi-metal.

The electric gyrotropic material half-space is characterized by the generic (relative) permittivity tensor $\bar{\varepsilon} = \varepsilon_t \mathbf{1}_t + \varepsilon_a \hat{\mathbf{z}} \otimes \hat{\mathbf{z}} + i\varepsilon_g \hat{\mathbf{z}} \times \mathbf{1}$ with $\mathbf{1}_t = \mathbf{1} - \hat{\mathbf{z}} \otimes \hat{\mathbf{z}}$. Then, the magnetic field in the frequency domain is of the form:

$$H_z(x,y,\omega) = \begin{cases} Ae^{-\gamma_0 y}e^{ik_x x}, & y > y_0 \\ B\left(e^{+\gamma_0 y} + Re^{-\gamma_0 y}\right)e^{ik_x x}, & 0 < y < y_0 \\ BTe^{\gamma_g y}e^{ik_x x}, & y < 0 \end{cases} \quad (1)$$

where $\gamma_0 = \sqrt{k_x^2 - \omega^2/c^2} = -i\sqrt{\omega^2/c^2 - k_x^2}$ is the free space propagation constant, $\gamma_g = \sqrt{k_x^2 - \varepsilon_{ef}\omega^2/c^2} = -i\sqrt{\varepsilon_{ef}\omega^2/c^2 - k_x^2}$ is the propagation constant in the gyrotropic material and $\varepsilon_{ef} = \dfrac{\varepsilon_{11}^2 + \varepsilon_{12}^2}{\varepsilon_{11}}$ is the effective permittivity of the gyrotropic medium. Note that $\varepsilon_{11} = \varepsilon_t$ and $\varepsilon_{12} = -i\varepsilon_g$. The coefficients $A$, $B$, $R$ and $T$ are determined by the boundary conditions. At the source plane $y = y_0$, the frequency domain magnetic field is discontinuous $H_z\big|_{y=y_0^+} - H_z\big|_{y=y_0^-} = -en_z \,\text{sgn}(v)e^{ik_x x}$, whereas $E_x$ remains continuous.

The latter condition is equivalent to $\left.\frac{\partial H_z}{\partial y}\right|_{y=y_0^+} - \left.\frac{\partial H_z}{\partial y}\right|_{y=y_0^-} = 0$. These two boundary conditions give:

$$A = -en_z \operatorname{sgn}(v) \frac{1}{2}\left(e^{+\gamma_0 y_0} - e^{-\gamma_0 y_0} R\right), \qquad B = en_z \operatorname{sgn}(v) \frac{e^{-\gamma_0 y_0}}{2}. \qquad (2)$$

Evidently, $R$ and $T$ can be identified as the reflection and transmission coefficients for plane wave incidence at a single interface between the vacuum and the gyrotropic material. Using mode matching one easily finds that [7]:

$$R = \frac{\dfrac{\gamma_0}{\varepsilon_0} - \left(\dfrac{\gamma_g}{\varepsilon_{ef}} + ik_x \dfrac{\varepsilon_{12}}{\varepsilon_{11}^2 + \varepsilon_{12}^2}\right)}{\dfrac{\gamma_0}{\varepsilon_0} + \left(\dfrac{\gamma_g}{\varepsilon_{ef}} + ik_x \dfrac{\varepsilon_{12}}{\varepsilon_{11}^2 + \varepsilon_{12}^2}\right)}, \qquad T = 1 + R. \qquad (3)$$

From Eq. (2), the frequency domain magnetic field may be written in a more compact manner as:

$$H_z(x,y,\omega) = \frac{1}{2} n_z e \operatorname{sgn}(v) h_z(x,y,\omega),$$
$$h_z(x,y,\omega) = \begin{cases} \left[\operatorname{sgn}(-y+y_0)e^{-\gamma_0|y-y_0|} + Re^{-\gamma_0(y+y_0)}\right]e^{ik_x x}, & y > 0 \\ e^{-\gamma_0 y_0} T e^{\gamma_g y} e^{ik_x x}, & y < 0 \end{cases}. \qquad (4)$$

Applying the inverse Fourier transform, one finds the time-domain magnetic field:

$$H_z(x,y,t) = \frac{1}{2} n_z e \operatorname{sgn}(v) \omega_p h_z(x,y,t)$$
$$h_z(x,y,t) = \frac{1}{2\pi} \int_{-\infty}^{+\infty} h_z(x,y,\omega) e^{-i\omega t} \frac{d\omega}{\omega_p}. \qquad (5)$$

The normalization parameter $\omega_p$ (with unities of frequency) was introduced so that the function $h_z(x,y,t)$ is dimensionless. In the numerical examples, $\omega_p$ will correspond to the plasma frequency of the relevant material. It is simple to prove that the dependence in time of the magnetic field is of the form $H_z(x,y,t) = H_z(x-vt, y, t=0)$. Hence, the emitted fields are characterized by a single snapshot in time.

The amount of energy extracted from the moving charges is the so-called "stopping power". Since the velocity of the charges is assumed time-independent, it may be written as $P_{ext} = -\int \mathbf{E}_{loc} \cdot \mathbf{j}_e \, d^3\mathbf{r}$, where $\mathbf{E}_{loc}$ represents the local electric field that acts on the charges and $\mathbf{j}_e(x,y,t)$ is the electric current density in the time domain. It can be written explicitly as:

$$\frac{P_{ext}}{L_z} = \frac{P_0}{y_0} \int_{-\infty}^{+\infty} G(\omega) \frac{d\omega}{\omega_p}, \tag{6}$$

where $P_0 = \frac{n_z^2 e^2}{4\pi\varepsilon_0} c$ is a normalization factor (with units of power), $L_z$ represents the width of the current pencil along the $z$-direction and,

$$G(\omega) = \text{Re}\left\{ \frac{|v|}{c} \gamma_0 y_0 \frac{\omega_p}{i\omega} R e^{-2\gamma_0 y_0} \right\} \tag{7}$$

is the (bilateral) normalized power spectral density that determines the spectrum of the emitted radiation.

### III. Cherenkov emission in the vicinity of a gyrotropic half-space

#### A. Magnetized plasma

For the particular case of a magnetized electron gas (e.g., a magnetized semiconductor [32]), the components of the permittivity tensor are [33]:

$$\varepsilon_t = 1 - \frac{\omega_p^2 (i\Gamma + \omega)}{\omega (i\Gamma + \omega)^2 - \omega\omega_0^2}, \quad \varepsilon_a = 1 - \frac{1}{\omega} \frac{\omega_p^2}{(i\Gamma + \omega)}, \quad \varepsilon_g = \frac{1}{\omega} \frac{\omega_0 \omega_p^2}{\omega_0^2 - (i\Gamma + \omega)^2}, \tag{8}$$

where $\omega_p$ is the plasma frequency, $\omega_0 = -qB_0/m$ is the cyclotron frequency ($q = -e$ is the negative charge of the electrons and $m$ is the effective mass), $\mathbf{B}_0 = B_0 \hat{\mathbf{z}}$ is the biasing magnetic field and $\Gamma$ is the collision frequency. Note that the theory of Sect. II neglects the influence of the magnetic field bias on the trajectory of the pencil beam. This is acceptable if the interaction of $\mathbf{B}_0$ with the moving charges is confined to a limited spatial region (with a characteristic dimension much smaller than the radius of the cyclotron orbit), which in principle is the case in a real experiment.

The band diagram of an unbounded (bulk) gyrotropic medium ($\omega = \omega_{k_x}^{\text{bulk}}$) is found from the solution of $k^2 = \varepsilon_{ef} \omega^2 / c^2$ (for propagation in the *xoy* plane, perpendicular to the bias field), and is plotted in Fig. 1b for the parameters $\omega_0 = 0.5\omega_p$ and $\Gamma = 0^+$. As seen, it exhibits a spectral symmetry such that $\omega_{k_x}^{\text{bulk}} = \omega_{-k_x}^{\text{bulk}}$.

The surface plasmons [see Eq. (9)] supported by the interface between the vacuum and the topological material half-space are determined by the dispersion equation [6]-[7], [34]-[36]

$$\frac{\gamma_0}{\varepsilon_0} + \frac{\gamma_g}{\varepsilon_{ef}} + ik_x \frac{\varepsilon_{12}}{\varepsilon_{11}^2 + \varepsilon_{12}^2} = 0 \ . \tag{9}$$

Figure 2a depicts the plasmons dispersion ($\omega = \omega_{k_x}^{\text{SPP}}$) for a lossless magnetized plasma with $\omega_0 = 0.5\omega_p$. In contrast to the bulk states, the SPPs dispersion is strongly asymmetric due to the inhomogeneity of space (which breaks the parity symmetry) and the nonreciprocity inherent to gyrotropic material. In particular, the plasmon resonance, i.e., the asymptotic value reached when $k_x \to \pm\infty$, depends on the direction of propagation of the plasmons. It is given by [37]

$$\lim_{k_x \to \pm\infty} \omega_{k_x}^{\text{SPP}} \equiv \omega_{\pm} = \frac{1}{2}\left(\mp\omega_0 + \sqrt{2\omega_p^2 + \omega_0^2}\right). \tag{10}$$

In the Cherenkov problem, the wave number along *x* and the frequency are linked by $k_x = \omega/v$. Hence, a natural (surface or bulk) mode of the system can be excited by the moving charges when the following selection rule is satisfied for some $k_x$:

$$\omega_{k_x} = k_x v. \tag{11}$$

Here, $\omega_{k_x}$ represents the dispersion of the relevant modes (e.g., for the SPP waves $\omega_{k_x} = \omega_{k_x}^{SPP}$). Figure 2a depicts the lines $\omega = \pm k_x v$ for two different values of the charges velocity. As seen, when $v > 0$ they intersect the plasmon dispersion at some point with $\omega \approx \omega_+$, whereas if $v < 0$ they intersect the plasmon dispersion at some point with $\omega \approx \omega_-$. Thus, the Cherenkov emission spectrum may be strongly asymmetric. For positive velocities it will be peaked near $\omega \approx \omega_+$, whereas for negative velocities it will be peaked near $\omega \approx \omega_-$. The moving charges can also be coupled to bulk modes with dispersion $\omega_{k_x}^{bulk}$. However, as previously discussed, $\omega_{k_x}^{bulk}$ is an even function of the wave number, and hence does not lead to any emission asymmetry.

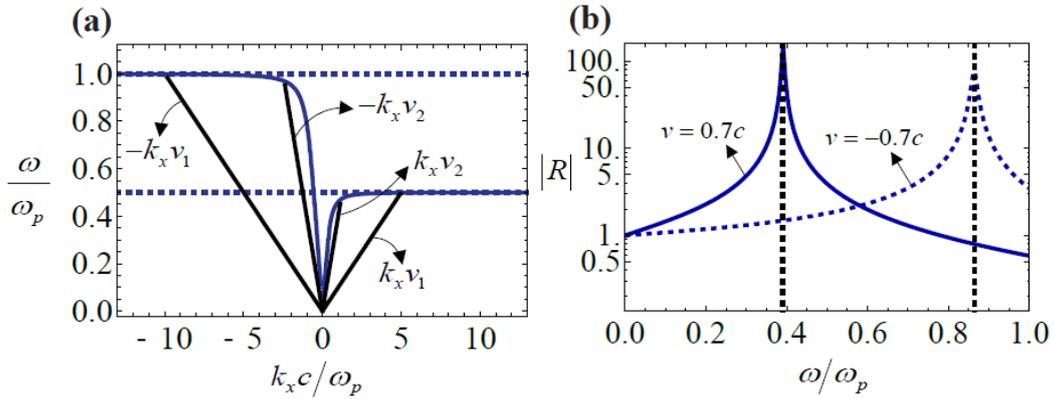

**Fig. 2 (a)** Dispersion diagrams of the edge (surface) plasmons propagating at an interface between air and a magnetized plasma with $\omega_0 = 0.5\omega_p$ (blue thick lines). The plasmon resonance occurs at $\omega = \omega_+$, for $k_x > 0$, and $\omega = \omega_-$ for $k_x < 0$. The thin black lines represent the curves $\omega = k_x v$ for the velocities $v_1 = \pm 0.1c$ and $v_2 = \pm 0.4c$. Their intersection with the dispersion diagram determines the main emission channels in the Cherenkov problem. **(b)** Amplitude of the reflection coefficient (in a logarithmic scale) for $v = \pm 0.7c$. The vertical gridlines correspond approximately to $\omega/\omega_p \approx 0.78\omega_+$ (curve $v = 0.7c$) and $\omega/\omega_p \approx 0.87\omega_-$ (curve $v = -0.7c$). The collision frequency was taken equal to $\Gamma = 0.01\omega_p$.

In the Cherenkov problem the amplitude of the scattered fields is determined by the reflection coefficient $R(\omega, k_x)\big|_{k_x = \omega/v}$. Figure 2b depicts the reflection coefficient amplitude as a function of frequency. Consistent with Fig. 2a, the reflection coefficients are highly asymmetric being peaked precisely at the frequencies at which the selection rule (11) is satisfied (vertical gridlines in Fig. 2b), i.e., relatively near $\omega_\pm$ depending on the sign of the electrons velocity.

The previous result suggests that the moving charges can efficiently excite the surface plasmons. Depending on the sign of the velocity, the spectrum of the emitted Cherenkov radiation will be peaked at the plasmon resonance $\omega_+$ or $\omega_-$. To demonstrate the plasmon excitation and the spectral asymmetry of the emitted radiation, we depict in Fig. 3 the frequency domain magnetic field [Eq. (4)] calculated at the frequencies $\omega = 0.94\omega_+$ [Figs. 3ai and bi], $\omega = 0.7\omega_-$ [Figs. 3aii and bii] and $\omega = 0.97\omega_-$ [Figs. 3aiii and biii]. The plots in Fig. 3a (Fig. 3b) correspond to positive (negative) values of the charges velocity. The dashed horizontal black line represents the trajectory of the charges and the solid horizontal black line the interface. As seen, the SPPs are strongly excited in the examples of Figs. 3ai [$\omega \approx \omega_+$ and $v > 0$] and 3biii [$\omega \approx \omega_-$ and $v < 0$] such that the magnetic field is concentrated near the interface $y=0$. For frequencies far from the plasmon resonances, $\omega_+$ or $\omega_-$ [Figs. 3aii and bii], the SPP emission is not observed, and the fields are concentrated near the trajectory of the moving charges.

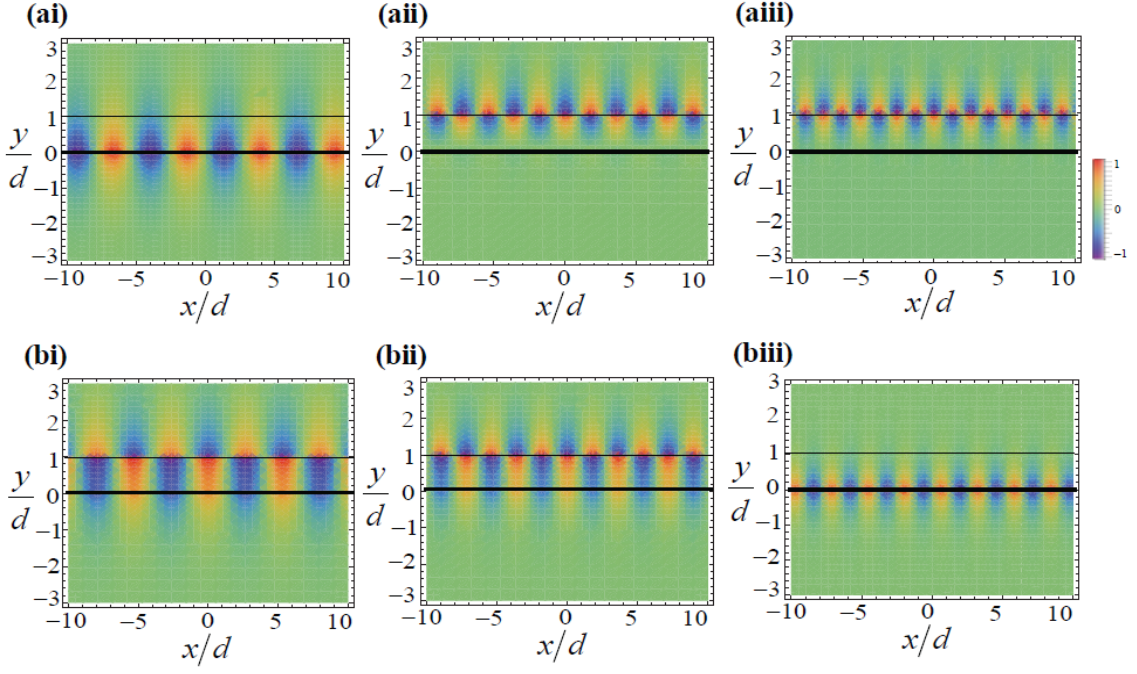

**Fig. 3** Real part of the magnetic field emission spectrum $H_z(x,y,\omega)$ (in arbitrary units) for an array of moving charges in the vicinity of a magnetized plasma with $\omega_0 = 0.5\omega_p$, $\omega_p d/c = 1$, and $\Gamma = 0^+$. **(a)** $v = 0.4c$ and (i) $\omega = 0.94\omega_+$, (ii) $\omega = 0.7\omega_-$, (iii) $\omega = 0.97\omega_-$. **(b)** Similar to (a) but for $v = -0.4c$.

Figure 4 shows a density plot of the instantaneous magnetic field emitted by the moving charges for different values of the velocity and material loss. The magnetic field was calculated using Eq. (6) for $t=0$, which corresponds to the charges position $x=0$. The time-domain simulation confirms that the SPP waves are indeed the main Cherenkov emission channels. Notably, a strongly asymmetric Cherenkov radiation is observed in the plots with opposite velocity signs [Figs. 4ai-aii and Figs. 4bi-bii]. In the case of stronger material absorption [Figs. 4aii and bii] the emitted SPPs are evidently damped after propagating a few guided wavelengths.

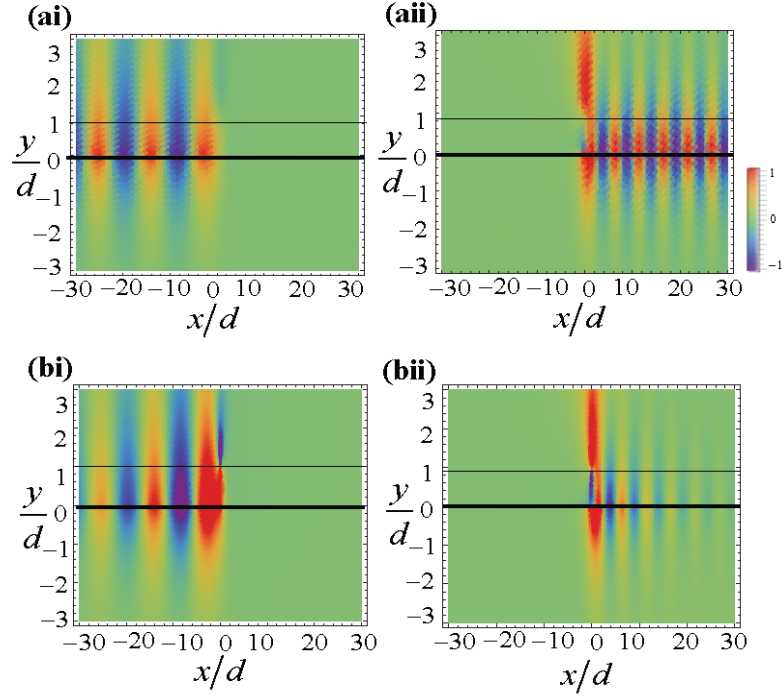

**Fig. 4** Time snapshot (*t=0*) of the magnetic field $H_z$ (in arbitrary units) for an array of moving charges in the vicinity of a magnetized plasma with cyclotron frequency $\omega_0 = 0.5\omega_p$ and for $\omega_p d/c = 1$. (**ai**) $v = 0.7c$, $\Gamma = 0.001\omega_p$ (**aii**) $v = -0.7c$, $\Gamma = 0.001\omega_p$. (**b**) Similar to (**a**) but with $\Gamma = 0.1\omega_p$.

Figure 5a depicts the normalized power spectral density $G$ [Eq. (7)] for $\Gamma = 0.01\omega_p$ and $\omega_0 = 0.5\omega_p$. Clearly, the emission spectrum is highly asymmetric being peaked near $\omega_\pm$ (vertical gridlines in Fig. 5a) depending on the sign of the electrons velocity. This property is explained by the fact that the density of (plasmon) states diverges at the frequencies determined by $\lim_{k_x \to \pm\infty} \omega_{k_x}^{SPP}$, and further underscores the relevance of plasmons in the Cherenkov problem. The "stopping power" [Eq. (6)] is represented in Fig. 5b for different values of distance *d*. For larger distances *d* the stopping power decreases because the coupling between the fields generated by the moving charges and the plasmons is weaker. As expected, the stopping power is typically larger for more energetic beams (with a larger *v*). However, after some velocity threshold value the stopping power drops sharply. The reason is that for large velocities the intersection of the line $\omega = k_x v$ with the plasmons dispersion $\omega_{k_x}^{SPP}$ occurs at points more distant from the "flat" parts of $\omega_{k_x}^{SPP}$ (see Fig. 2a), where the density of states is smaller.

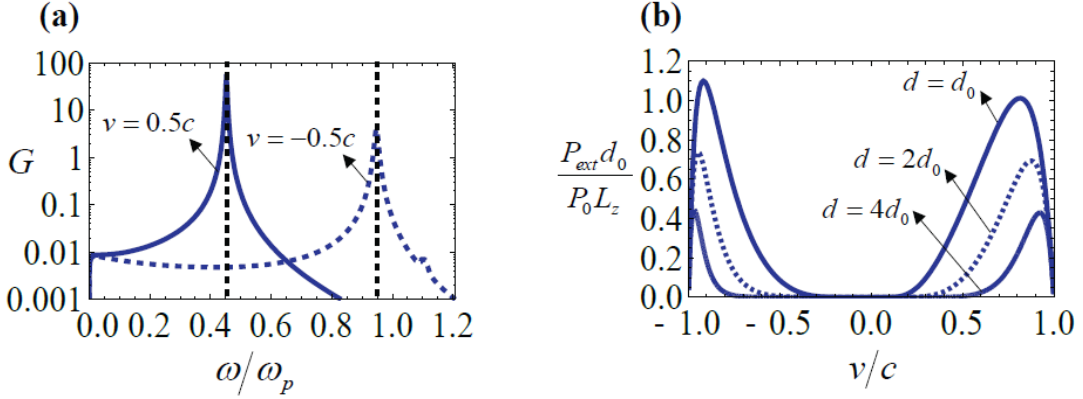

**Fig. 5** (a) Normalized power spectral density $G$ as a function of the normalized frequency for $\omega_p d / c = 1$. The vertical gridlines indicate the points that satisfy the selection rule (11), relatively near the plasmon resonances $\omega \approx \omega_\pm$. (b) Normalized stopping power $P_{ext}$ as a function of the charge velocity $v$, for $d = d_0$, $d = 2d_0$ and $d = 4d_0$ with $\omega_p d_0 / c = 1$.

### B. Weyl semi-metal

As a second example, we consider a linear array of charges moving in a vacuum above a Weyl semi-metal half-space. It has been suggested that pyrochlore iridates of the generic form $A_2Ir_2O_7$, with A either yttrium or a lanthanide element, may be topological semi-metals [15-16]. The topological properties of a Weyl semi-metal results in an anomalous Hall current (chiral anomaly) and creates a gyrotropic (nonreciprocal) electromagnetic response, even *without* an external magnetic bias [17]. Thus, Weyl semi-metals may have a spontaneous nonreciprocal response due to magnetic ordering [15]. There are other naturally available materials with a spontaneous nonreciprocal response, e.g., some antiferromagnets [38]-[43].

Following Ref. [17], a Weyl semi-metal has a permittivity tensor of the form $\overline{\varepsilon} = \varepsilon_t \mathbf{1} + i\varepsilon_g \hat{\mathbf{z}} \times \mathbf{1}$ with permittivity components:

$$\varepsilon_t = \varepsilon_\infty \left(1 - \frac{\omega_p^2}{\omega^2}\right), \qquad \varepsilon_g = \varepsilon_\infty \frac{\omega_b}{\omega}, \qquad (12)$$

where $\varepsilon_\infty$ is the high-frequency permittivity, $\omega_p$ denotes the bulk plasma frequency and $\omega_b$ is a parameter with units of frequency which is nontrival for topological Weyl semi-metals [17]. Here, the z-axis is taken as the direction of the wave vector **b** that links the two Weyl points in the 1$^{st}$ Brillouin zone [17]. Figure 1c shows the band diagram of a bulk Weyl semi-metal for the case $\varepsilon_\infty = 1$ and $\omega_b = 0.5\omega_p$. Similar to the magnetized plasma case, the band diagram exhibits spectral symmetry.

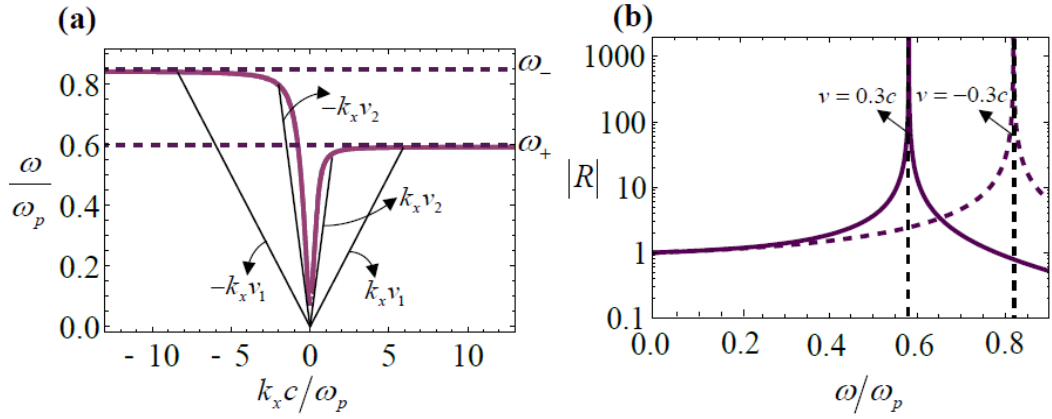

**Fig. 6 (a)** Dispersion diagrams of the edge (surface) plasmons propagating at an interface between air and a Weyl semi-metal (purple thick lines). The plasmon resonance occurs at $\omega_+ \approx 0.59\omega_p$, for $k_x > 0$, and $\omega_- \approx 0.84\omega_p$, for $k_x < 0$. The thin black lines represent the curves $\omega = k_x v$ for the velocities $v_1 = \pm 0.1c$ and $v_2 = \pm 0.4c$. **(b)** Amplitude of the reflection coefficient (in a logarithmic scale) for $v = \pm 0.3c$. The vertical gridlines correspond approximately to $\omega \approx 0.99\omega_+$ (curve $v = 0.3c$) and $\omega \approx 0.98\omega_-$ (curve $v = -0.3c$).

The plasmons dispersion diagram is plotted in Fig. 6a for the parameters $\varepsilon_\infty = 1$ and $\omega_b = 0.5\omega_p$ showing a behavior qualitatively similar to that of the magnetized plasma case. Now, the dispersion curves saturate at the plasmon resonances:

$$\lim_{k_x \to \pm\infty} \omega_{k_x}^{SPP} \equiv \omega_{\pm} = \frac{\varepsilon_{\infty}}{2(1+\varepsilon_{\infty})}\left(\mp \omega_b + \sqrt{\frac{1+\varepsilon_{\infty}}{\varepsilon_{\infty}}4\omega_p^2 + \omega_b^2}\right). \qquad (13)$$

Similar to Fig. 2b, the reflection coefficient is typically peaked near the plasmon resonances $\omega_{\pm}$ (see Fig. 6b).

Furthermore, the magnetic field emission spectrum (not shown) and the instantaneous magnetic field (Fig. 7) have features qualitatively very similar to the magnetized plasma examples. They confirm the selective emission of plasmons with spectrum concentrated near $\omega_{\pm}$, depending if the pencil beam travels along the +x or -x-direction, and that plasmons are typically the main emission channels.

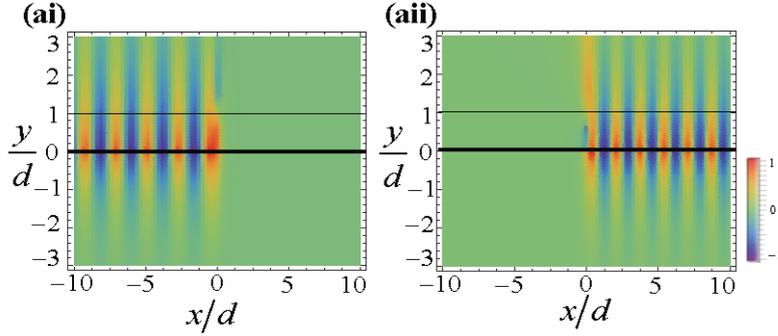

**Fig. 7** Time snapshot ($t=0$) of the magnetic field $H_z$ (in arbitrary units) for an array of moving charges in the vicinity of a Weyl semi-metal with $\varepsilon_{\infty}=1$, $\omega_b = 0.5\omega_p$ and $\omega_p d/c = 1$. (**ai**) $v=0.7c$, (**aii**) $v=-0.7c$.

The power spectral density $G$ [Fig. 8a] and the stopping power [Fig. 8b] also have features qualitatively similar to the magnetized plasma case. For example, the power spectral density is resonant near $\omega_{\pm}$, depending on the sign of $v$. However, it must be highlighted that the nonreciprocal response of the Weyl semi-metal is spontaneous and hence it does not require an external magnetic bias.

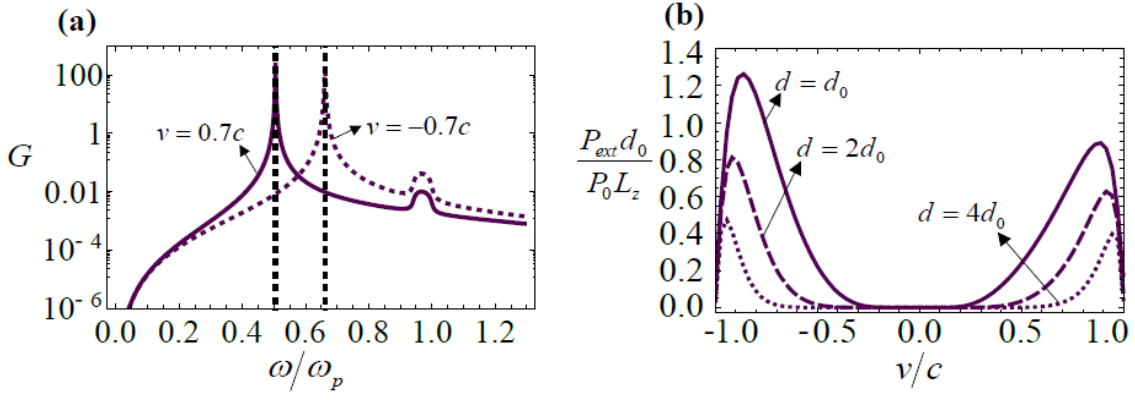

**Fig. 8** (a) Normalized power spectral density $G$ as a function of the normalized frequency for $\omega_p d / c = 1$. The vertical gridlines indicate the points that satisfy the selection rule (11). (b) Normalized stopping power $P_{ext}$ as a function of the charge velocity $v$, for $d = d_0$, $d = 2d_0$ and $d = 4d_0$ with $\omega_p d_0 / c = 1$. In the simulations we used $\varepsilon_\infty = 1$ and $\omega_b = 0.5\omega_p$.

In the previous examples, it was assumed that $\varepsilon_\infty = 1$ but it may be more realistic to take $\varepsilon_\infty > 1$ in the material response model [17]. The reason why we used $\varepsilon_\infty = 1$ is that in the Cherenkov problem the high-frequency response is of a crucial importance. For example, taking $\varepsilon_\infty \neq 1$ when the charged particles move inside a material leads to divergences and to an infinite stopping power [26]. Hence, usually it is essential to consider realistic dispersive models that fully satisfy the Kramers-Kronig relations with $\varepsilon_\infty = 1$ [20]. In our case, the beam of charges moves outside the material and in this situation the distance $d$ to the material provides a natural cut-off, which avoids the aforementioned divergences. However, taking $\varepsilon_\infty \neq 1$ may overestimate the amount of radiation emitted in the form of bulk states. With this note of caution, we represent in Figure 9 the instantaneous magnetic field calculated for different values of $\varepsilon_\infty$. As seen, for increasingly larger values of $\varepsilon_\infty$ the amount of radiation emitted in the form of bulk states increases, and the overall importance of the surfaces plasmons decreases. Indeed, a well-defined Cherenkov cone becomes visible in the region $y < 0$ when $\varepsilon_\infty \neq 1$. The cone becomes more tilted as $\varepsilon_\infty$ increases. Evidently, even when $\varepsilon_\infty \neq 1$ the surface plasmons remain the main emission channels in the spectral windows near $\omega_\pm$.

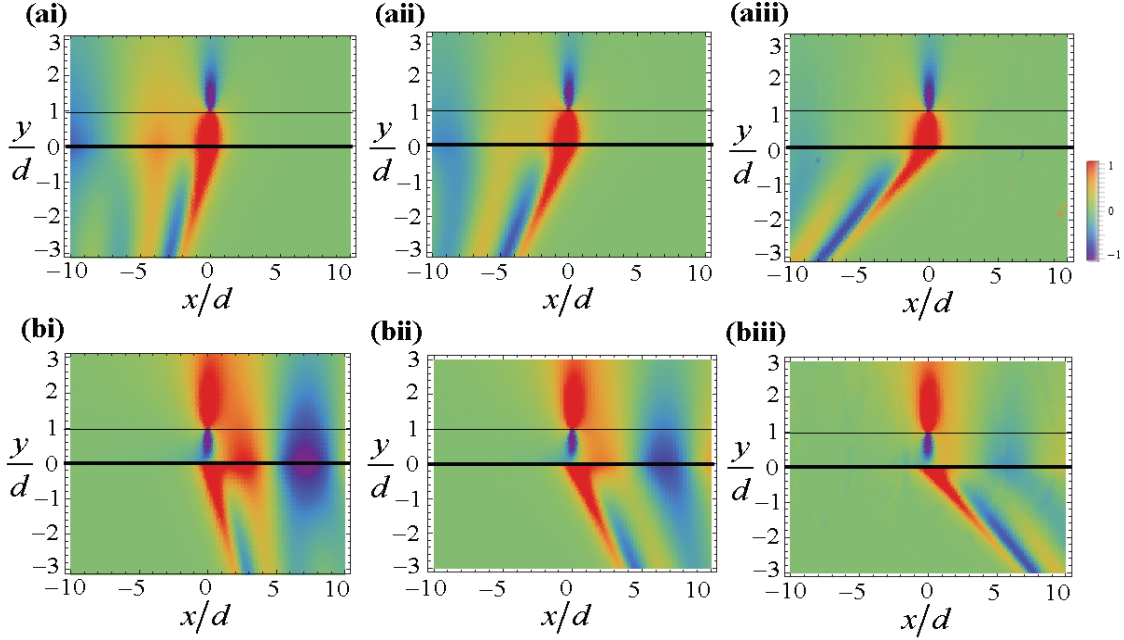

**Fig. 9** Time snapshot ($t=0$) of the magnetic field $H_z(x,y,t)$ (in arbitrary units) for a linear array of moving charges for $\omega_b = 0.5\omega_p$ and $\omega_p d/c = 1$. **(a)** The charge velocity is $v = 0.7c$ and the high-frequency permittivity is (i) $\varepsilon_\infty = 3$ (ii) $\varepsilon_\infty = 5$ (iii) $\varepsilon_\infty = 13$. **(b)** Similar to (a) but with $v = -0.7c$.

## IV. Analytical solution for the stopping power

If the charges velocity satisfies $|v| \ll c$, and, if in addition the electrons move sufficiently close to the surface of the topological material ($\omega_p d/c \ll 1$), the effects of time retardation can be neglected. In such a case, the light-matter interactions are essentially quasi-static and in particular it is possible to characterize the power extracted from the moving beam using a quasi-static approximation.

To this end, we rely on the formalism of Appendix A which gives the stopping power in terms of the electromagnetic modes of the system [Eq. (A10)]. In the quasi-static limit, the main radiation channels are determined by the surface plasmons. For simplicity, we restrict the analysis to the case of a gyrotropic material with dispersion as in Eq. (8). It was recently shown that the (short-wavelength) surface plasmons at an air-gyrotropic material interface are of the form $\mathbf{E}_\mathbf{k} \approx -\nabla \phi_\mathbf{k}$, $\mathbf{H}_\mathbf{k} \approx 0$ with,

$$\phi_\mathbf{k} = A_\mathbf{k} e^{i\mathbf{k}\cdot\mathbf{r}} \begin{cases} e^{-k_\| y}, & y > 0 \\ e^{+\tilde{k}_\| y}, & y < 0 \end{cases}, \tag{14}$$

with $\mathbf{k} = (k_x, 0, k_z)$ the transverse wave vector (parallel to the interface), $k_\| = \sqrt{k_x^2 + k_z^2}$, $\tilde{k}_\| = \sqrt{k_x^2 + \frac{\varepsilon_a}{\varepsilon_t} k_z^2}$ and $A_\mathbf{k}$ a normalization constant. Furthermore, the dispersion of the short-wavelength plasmons satisfies (the formulas of this article differ slightly from those of Ref. [37] due to a different choice of the coordinate axes) [37]

$$\omega_\mathbf{k} \equiv \omega_\varphi = -\frac{\omega_0}{2}\cos\varphi + \sqrt{\frac{\omega_p^2}{2} + \frac{\omega_0^2}{4}(1+\sin^2\varphi)}, \tag{15}$$

with $\varphi$ the angle of the SPP wave vector $\mathbf{k}$ with respect to the $x$-axis. As discussed in the detail in Ref. [37], one has $\lim_{k\to+\infty} \omega^{SPP}_{k(\cos\varphi, 0, \sin\varphi)} \equiv \omega_\varphi$ and in particular $\omega_{\varphi=0,\pi} = \omega_\pm$. Thus, the quasi-static approximation describes the plasmon resonances with a large wave vector. The plasmon resonances are direction dependent due to the magnetic bias.

In Appendix B, we use the quasi-static model to determine the stopping power. For the case of a pencil beam it is found that:

$$\frac{P_{ext}}{L_z} = P_0 \frac{4\pi \dfrac{\omega_\pm}{c} e^{-2\frac{\omega_\pm y_0}{v}}}{\left[1 + \partial_\omega(\varepsilon_t \omega) - \partial_\omega(\varepsilon_g \omega)\mathrm{sgn}(v)\right]_{\omega=\omega_\pm}}, \tag{16}$$

with $P_0 = \dfrac{n_z^2 e^2}{4\pi\varepsilon_0} c$. The "+" sign ("−" sign) is chosen for positive (negative) velocities. A comparison between this quasi-static result and the exact theory of Sect. II is reported in Fig. 10a. As seen, the agreement is excellent for velocities $|v|/c < 0.3$. For large velocities the quasi-static approximation breaks down because the effects of time-retardation become relevant (as previously discussed, for large velocities the excited plasmons have long wavelengths, i.e. a small **k**; hence they cannot be modeled by Eq. (15), which describes the dispersion of the short-wavelength plasmons).

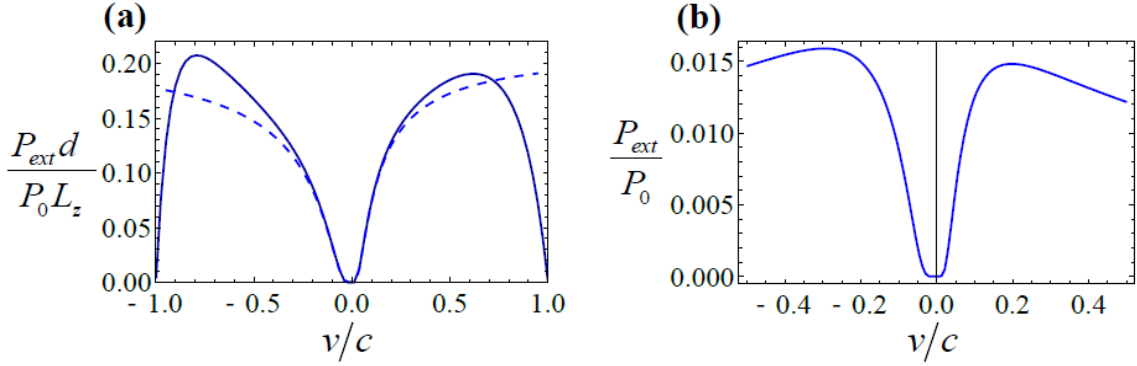

**Fig. 10 (a)** Comparison between the exact stopping power (solid line) and the stopping power obtained with the quasi-static approximation [Eq. (16)] (dashed line) for a "pencil" beam. **(b)** Stopping power obtained with the quasi-static approximation [Eq. (17)] for a "point-like" beam. In both plots we use $\omega_p d/c = 0.1$ and $\omega_0/\omega_p = 0.4$.

Furthermore, in Appendix B we also determine the stopping power for the case of a "point-like" beam described by the current density $\mathbf{j}_e(x,y,t) = -ev\delta(z)\delta(y-y_0)\delta(x-vt)\hat{\mathbf{x}}$ (this current density corresponds to a single moving electron). The quasi-static solution of this fully three-dimensional radiation problem is:

$$P_{ext} = P_0 \int_{v\cos\varphi > 0} d\varphi \frac{\omega_\varphi^2 y_0^2}{|cv\cos\varphi|} a_\varphi e^{-2\frac{\omega_\varphi}{v\cos\varphi}y_0} . \qquad (17)$$

where the normalization factor is now $P_0 = \frac{e^2 c}{4\pi\varepsilon_0}\frac{1}{y_0^2}$ and $y_0 = d$. In the above, $\omega_\varphi$ is defined by Eq. (15) and $a_\varphi = \varepsilon_0 k_\parallel |A_\mathbf{k}|^2$ with $|A_\mathbf{k}|^2$ given by Eq. (B3). The integration range is over $-\pi/2 < \varphi < \pi/2$ for positive velocities and over $\pi/2 < \varphi < 3\pi/2$ for negative velocities. The computed stopping power is shown in Fig. 10b as a function of the charge velocity. As seen, the results are qualitatively similar to the pencil beam case and exhibit a pronounced asymmetry such that the stopping power depends strongly on the sign of the velocity.

## V. Conclusions

In this work it was demonstrated that the main radiation channels of the Cherenkov energy emitted by an electron beam moving near to a topological material half-space are the edge states (plasmons) supported by the interface. It was shown that the emission spectrum is highly asymmetric, being peaked near a plasmon resonance that depends on the sign of the electrons velocity. The theory was applied to the cases of a magnetized plasma and Weyl semimetals. In particular, it was found that Weyl semimetals offer the opportunity to obtain strongly asymmetric Cherenkov emissions with no biasing magnetic field, in contrast to magnetized plasmas. The platforms described in this work offer the possibility to discriminate particles travelling the opposite directions of space in a simple manner, which can have applications in particle detection and biomedicine.

**Acknowledgements:** This work is supported in part by Fundação para a Ciência e a Tecnologia grant number PTDC/EEI-TEL/4543/2014 and by Instituto de Telecomunicações under project UID/EEA/50008/2013. F. R. Prudêncio acknowledges financial support by Fundação para a Ciência e a Tecnologia (FCT) under the Post-Doctoral fellowship SFRH/BPD/108823/2015.

## Appendix A: Modal expansion of the radiated fields and stopping power

In this Appendix, we obtain explicit formulas for the fields emitted by a moving charge distribution and for the stopping power in terms of a modal expansion [27, 44, 45]. The derived formulas are exact and the only assumption is that the material loss is negligible.

To begin with, we determine the fields emitted by a generic current density. For simplicity, we use six-vector notations and denote the electromagnetic fields as $\mathbf{f}(\mathbf{r},t) = (\mathbf{E} \quad \mathbf{H})^T$ and the excitation currents as $\mathbf{j}(\mathbf{r},t) = (\mathbf{j}_e \quad \mathbf{j}_m)^T$. Here, $\mathbf{j}_e$ is the instantaneous electric current density and $\mathbf{j}_m$ is the instantaneous magnetic current density, which will be taken identical to zero. Furthermore, we introduce a 6×6 frequency-domain Green tensor (with a time-harmonic variation $e^{-i\omega t}$) defined as the solution of [44, 45]:

$$\hat{N} \cdot \overline{\mathbf{G}}(\mathbf{r},\mathbf{r}_0,\omega) = \omega \mathbf{M}(\mathbf{r},\omega) \cdot \overline{\mathbf{G}}(\mathbf{r},\mathbf{r}_0,\omega) + i\mathbf{1}\delta(\mathbf{r} - \mathbf{r}_0). \tag{A1}$$

with $\hat{N} = \begin{pmatrix} \mathbf{0} & i\nabla \times \mathbf{1}_{3\times 3} \\ -i\nabla \times \mathbf{1}_{3\times 3} & \mathbf{0} \end{pmatrix}$, $\mathbf{1}$ the identity matrix of dimension six, and

$\mathbf{M} = \begin{pmatrix} \varepsilon_0 \overline{\varepsilon}(\mathbf{r},\omega) & \mathbf{0} \\ \mathbf{0} & \mu_0 \mathbf{1}_{3\times 3} \end{pmatrix}$ the material matrix that determines the electromagnetic response of the relevant materials.

Let $\mathbf{j}_\omega(\mathbf{r})$ be the Fourier transform of the generalized current vector $\mathbf{j}(\mathbf{r},t)$. Then, the Fourier transform of the fields satisfies:

$$\mathbf{f}_\omega(\mathbf{r}) = \int d^3\mathbf{r}' \overline{\mathbf{G}}(\mathbf{r},\mathbf{r}',\omega) \cdot \mathbf{j}_\omega(\mathbf{r}'). \tag{A2}$$

In the limit of vanishing material loss, the Green function has the modal expansion [44, 45]:

$$\overline{\mathbf{G}} = \frac{i}{2} \sum_n \frac{1}{\omega_n - \omega} \mathbf{f}_n(\mathbf{r}) \otimes \mathbf{f}_n^*(\mathbf{r}_0). \tag{A3}$$

Here, $\mathbf{f}_n(\mathbf{r})$ represents a generic electromagnetic mode with oscillation frequency $\omega_n$ and is normalized as (V stands for the volume of the relevant region)

$$\frac{1}{2}\int_V d^3\mathbf{r}\, \mathbf{f}_n^*(\mathbf{r})\cdot\frac{\partial}{\partial\omega}\left[\omega\mathbf{M}(\mathbf{r},\omega)\right]_{\omega=\omega_n}\cdot\mathbf{f}_n(\mathbf{r})=1.$$ The summation is over all the electromagnetic modes with either positive, zero or negative frequencies [44, 45].

For the geometry of interest (Fig. 1a), the modes are of the form $\mathbf{f}_n(\mathbf{r})\to\frac{1}{\sqrt{A_{tot}}}\mathbf{F}_{n\mathbf{q}}(y)e^{i\mathbf{q}\cdot\mathbf{r}}$ with $\mathbf{q}=(q_x,0,q_z)$ the transverse wave vector (which is a good quantum number due to the translational invariance along the $x$ and $z$ directions) and $A_{tot}$ the transverse area (section parallel to the interface). Taking into account that the modes form a continuum ($A_{tot}\to\infty$), it follows that:

$$\overline{\mathbf{G}}=\frac{i}{2}\sum_n\frac{1}{(2\pi)^2}\int d^2\mathbf{q}\frac{1}{\omega_{n\mathbf{q}}-\omega}\mathbf{F}_{n\mathbf{q}}(y)\otimes\mathbf{F}_{n\mathbf{q}}^*(y_0)e^{i\mathbf{q}\cdot(\mathbf{r}-\mathbf{r}_0)}. \tag{A4}$$

with $d^2\mathbf{q}=dq_x dq_z$ and the envelopes $\mathbf{F}_{n\mathbf{q}}$ normalized as:

$$\frac{1}{2}\int dy\,\mathbf{F}_{n\mathbf{q}}^*(y)\cdot\frac{\partial}{\partial\omega}\left[\omega\mathbf{M}(y,\omega)\right]_{\omega=\omega_{n\mathbf{q}}}\cdot\mathbf{F}_{n\mathbf{q}}(y)=1. \tag{A5}$$

Hence, the radiated field in the spectral domain is:

$$\mathbf{f}_\omega(\mathbf{r})=\frac{i}{2}\sum_n\frac{1}{(2\pi)^2}\int d^2\mathbf{q}\frac{1}{\omega_{n\mathbf{q}}-\omega}\mathbf{F}_{n\mathbf{q}}(y)\int d^3\mathbf{r}'\mathbf{F}_{n\mathbf{q}}^*(y')e^{i\mathbf{q}\cdot(\mathbf{r}-\mathbf{r}')}\cdot\mathbf{j}_\omega(\mathbf{r}'). \tag{A6}$$

Let us now suppose that the time-dependence of the current is of the form $\mathbf{j}=\mathbf{j}_0(y,z)\delta(x-vt)$ as in the Cherenkov problem. Then, $\mathbf{j}_\omega=\frac{1}{|v|}\mathbf{j}_0(y,z)e^{ik_x x}$ where $k_x=\omega/v$. From here, we can write (after carrying out the integration in $x'$):

$$\mathbf{f}_\omega(\mathbf{r})=\frac{i}{2|v|}\sum_n\frac{1}{2\pi}\int d^2\mathbf{q}\,e^{iq_x x}\delta\left(q_x-\frac{\omega}{v}\right)\frac{1}{\omega_{n\mathbf{q}}-\omega}\mathbf{F}_{n\mathbf{q}}(y)\int dy'dz'\mathbf{F}_{n\mathbf{q}}^*(y')\cdot\mathbf{j}_0(y',z')e^{iq_z(z-z')}.$$

(A7)

Calculating the inverse Fourier transform in time $\mathbf{f} = \frac{1}{2\pi}\int d\omega \mathbf{f}_\omega e^{-i\omega t}$ (the integration contour is slightly above the real-frequency axis), we obtain the following modal expansion for the radiated field:

$$\mathbf{f}(\mathbf{r},t) = \frac{i}{2}\sum_n \frac{1}{(2\pi)^2}\int d^2\mathbf{q}\, e^{iq_x(x-vt)} \frac{1}{\omega_{n\mathbf{q}} - q_x v - i0^+} \mathbf{F}_{n\mathbf{q}}(y)\int dy'dz'\, \mathbf{F}^*_{n\mathbf{q}}(y')\cdot \mathbf{j}_0(y',z') e^{iq_z(z-z')}.$$

(A8)

From the above result, one may write the stopping power ($P_{\text{ext}} = -\int dV \mathbf{j}\cdot\mathbf{f}$) in terms of the electromagnetic modes. Using again $\mathbf{j} = \mathbf{j}_0(y,z)\delta(x-vt)$ and noting that $\mathbf{j}_0(y,z)$ is real-valued, one obtains:

$$P_{\text{ext}} = \frac{-i}{2}\sum_n \frac{1}{(2\pi)^2}\int d^2\mathbf{q}\, \frac{1}{\omega_{n\mathbf{q}} - q_x v - i0^+} \left|\int dydz\, e^{iq_z z}\mathbf{F}_{n\mathbf{q}}(y)\cdot\mathbf{j}_0(y,z)\right|^2. \quad (A9)$$

Clearly, the stopping power must a real-valued number. On the other hand, the integrand is pure imaginary except for the contributions of the $i0^+$ term near the poles. Using $\frac{1}{x-i0^+} = \text{PV}\frac{1}{x} + i\pi\delta(x)$ (PV stands for the "principal value") we find that

$$P_{\text{ext}} = \frac{1}{4\pi}\sum_{\omega_{n\mathbf{q}}>0}\int d^2\mathbf{q}\,\delta(\omega_{n\mathbf{q}} - q_x v)\left|\int dydz\, e^{iq_z z}\mathbf{F}_{n\mathbf{q}}(y)\cdot\mathbf{j}_0(y,z)\right|^2. \quad (A10)$$

Consistent with the selection rule (11) only modes that satisfy $\omega_{n\mathbf{q}} = q_x v$ contribute to the stopping power. The summation was restricted to positive frequency modes because the modes with $\omega_{n\mathbf{q}} < 0$ give the same contribution as the modes with $\omega_{n\mathbf{q}} > 0$.

## Appendix B: Stopping power in a quasi-static approximation

In this Appendix, we derive explicit analytical formulas for the stopping power using the quasi-static approximation discussed in Sect. IV. We analyze both the cases of a pencil beam and of a point-like beam.

### I. Pencil beam

We consider first a pencil beam with $\mathbf{j}_0(y,z) = -e n_z v \delta(y-y_0)\hat{\mathbf{u}}_1$ with $\hat{\mathbf{u}}_1 = \begin{pmatrix} \hat{\mathbf{x}} \\ \mathbf{0} \end{pmatrix}$. Within the quasi-static approximation there is a single branch of positive-frequency modes and hence the summation over $n$ can be dropped in Eq. (A10). Denoting the wave vector of the surface plasmons with the symbol $\mathbf{k}$ ($n\mathbf{q} \to \mathbf{k}$), it follows that the stopping power is:

$$\frac{P_{\text{ext}}}{L_z} = \frac{(evn_z)^2}{2} \int d^2\mathbf{k}\, \delta(\omega_\mathbf{k} - k_x v)\delta(k_z) \left|\mathbf{F}_\mathbf{k}(y_0) \cdot \hat{\mathbf{u}}_1\right|^2. \tag{B1}$$

Furthermore, from Eq. (15) one has $\omega_\mathbf{k} = \omega_+$ for $k_x > 0$ and $k_z = 0$ and $\omega_\mathbf{k} = \omega_-$ for $k_x < 0$ and $k_z = 0$. This observation leads to:

$$\frac{P_{\text{ext}}}{L_z} = \frac{(en_z)^2 |v|}{2} \left|\mathbf{F}_\mathbf{k}(y_0) \cdot \hat{\mathbf{u}}_1\right|^2_{\substack{k_x = \omega_\pm/v \\ k_z = 0}}. \tag{B2}$$

where the "+" sign ("−" sign) is chosen for positive (negative) velocities.

To proceed, we use the fact that within the quasi-static approximation the fields have an electrostatic nature such that $\mathbf{F}_\mathbf{k} \approx \begin{pmatrix} \mathbf{E}_\mathbf{k} \\ 0 \end{pmatrix} \approx \begin{pmatrix} -\nabla\phi_\mathbf{k} \\ 0 \end{pmatrix}$ [37], with $\phi_\mathbf{k}$ given by Eq. (14). The constant $A_\mathbf{k}$ is determined by the normalization condition (A5). It can be written explicitly as (the difference compared to Ref. [37] is again due to the different coordinate system) [37]:

$$|A_{\mathbf{k}}|^2 = \frac{2}{\varepsilon_0}\left[k_{\|} + \frac{1}{2\tilde{k}_{\|}}\left(\partial_\omega(\varepsilon_t\omega)(\tilde{k}_{\|}^2 + k_x^2) + \partial_\omega(\varepsilon_a\omega)k_z^2 - \partial_\omega(\varepsilon_g\omega)2k_x\tilde{k}_{\|}\right)\right]^{-1}. \quad (B3)$$

Using these results to simplify (B2), one obtains:

$$\frac{P_{\text{ext}}}{L_z} = \frac{(en_z)^2}{\varepsilon_0}\frac{\omega_\pm e^{-2\frac{\omega_\pm y_0}{v}}}{\left[1 + \partial_\omega(\varepsilon_t\omega) - \partial_\omega(\varepsilon_g\omega)\text{sgn}(v)\right]_{\omega=\omega_\pm}}. \quad (B4)$$

In particular, the stopping power may be written as in Eq. (16) of the main text.

### *II. Point-like beam*

For a point-like beam, we have $\mathbf{j}_0(y,z) = -ev\delta(z)\delta(y-y_0)\hat{\mathbf{u}}_1$. Therefore, applying the quasi-static approximation to Eq. (A10), it is found that:

$$P_{\text{ext}} = \frac{(ev)^2}{4\pi}\int d^2\mathbf{k}\,\delta(\omega_{\mathbf{k}} - k_x v)|\mathbf{E}_{\mathbf{k}}(y_0)\cdot\hat{\mathbf{x}}|^2. \quad (B5)$$

Using $\mathbf{E}_{\mathbf{k}} \approx -\nabla\phi_{\mathbf{k}}$ and Eq. (14) it follows that:

$$\begin{aligned}P_{\text{ext}} &= \frac{e^2v^2}{4\pi}\int d^2\mathbf{k}\,\delta(\omega_\varphi - k_x v)k_x^2|A_{\mathbf{k}}|^2 e^{-2k_{\|}y_0} \\ &= \frac{e^2v^2}{4\pi\varepsilon_0}\int_{v\cos\varphi>0} d\varphi\frac{1}{|v\cos\varphi|}(k_{\|}\cos\varphi)^2 a_\varphi e^{-2k_{\|}y_0}\bigg|_{k_{\|} = \frac{\omega_\varphi}{v\cos\varphi}}\end{aligned}. \quad (B6)$$

In the second identity, we switched to polar coordinates ( $(k_x, k_z) = k_{\|}(\cos\varphi, \sin\varphi)$ ) and introduced $a_\varphi = \varepsilon_0 k_{\|}|A_{\mathbf{k}}|^2$, which only depends on the angle of the transverse wave vector, not on the amplitude. From the above result one readily obtains Eq. (17) of the main text.